\documentclass[a4paper,11pt]{article}

\usepackage{amsmath}
\usepackage{graphicx}
\usepackage{dcolumn}
\usepackage{bm}

\begin{document}

\hfill ZAGREB--ZTF--03/03

\begin{center}
\baselineskip=2\baselineskip
\textbf{\huge{
On Distinguishing Non-Standard Interactions from Radiative
Corrections in Neutrino--Electron Scattering}}\\[4ex]
\baselineskip=0.5\baselineskip


{\Large Kre\v{s}imir Kumeri\v{c}ki\footnote{kkumer@phy.hr} and
Ivica Picek\footnote{picek@phy.hr}}\\[2ex]
Department of Physics, Faculty of Science, University of Zagreb,
 P.O.B. 331, HR-10002 Zagreb, Croatia\\[3ex]
\today\\[5ex]
\end{center}

\begin{abstract}
We present a contribution of higher order to neutrino--electron
scattering that is a charged-current counterpart of both the anomalous
axial-vector triangle and possible non-standard interaction contributions.
It arises in 
the standard model with massive neutrinos, and renormalizes
the nondiagonal axial-vector form-factor at low energies.
We show that, due to the small size of radiative corrections, the
neutrino--electron scattering still provides a discovery potential
for some of the non-standard neutrino interactions proposed in the literature.
\end{abstract}

\vspace*{1ex}
\begin{flushleft}
\small
PACS: 11.15.-q, 13.40.Ks, 14.60.Pq
\vspace*{-2ex}
\begin{tabbing}
Keywords: \= neutrino conversion, non-standard interactions of neutrinos,\\
         \> radiative corrections, flavour mixing
\end{tabbing}
\end{flushleft}

\section{Introduction}

We have reached the time where the advance in neutrino experiments
enables quite accurate global fits on the leptonic mixing angles.
Simultaneously, the advent of the neutrino masses gives hope to infer 
on the specific departures from
the standard model (SM). In this sense we explore here a further
discovery potential of the neutrino--electron scattering.

 Ever since the earliest studies of the neutrino--electron scattering,
this particular process remained in a focus of interest.
The underlying \emph{diagonal} $\nu$--$e$ interaction
density is provided by the SM, which at energies much lower 
than $M_W$ gives
\begin{equation}
\mathcal{H}^{\text{diag}}(x)=\frac{G_F}{\sqrt{2}}\,\bar{e}\gamma^\mu 
 (g_V - g_A\gamma_5)
e\, \bar{\nu}\gamma_{\mu}(1-\gamma_5)\nu \;.
\label{ham}
\end{equation}
Here $g_V$ and $g_A$ are the SM couplings, where both
charged and neutral current contribute for $\nu=\nu_e\,$ 
($g_V  =  1/2+2\sin^{2}\theta_{W}$, $g_A=1/2$), 
whereas there is only neutral current contribution for 
$\nu= \nu_\mu, \nu_\tau$ 
($g_V  =  -1/2+2\sin^{2}\theta_{W}$, $g_A=-1/2$).

On the experimental side,
the results for atmospheric neutrinos \cite{SK98}, in conjunction 
with the results from the Sudbury
Neutrino Observatory (SNO) \cite{SNO01}, show
that the disappearance of one neutrino flavour (such as reported by
the SuperKamiokande)
is accompanied by the appearance of another.  Such conversion of the
neutrino species appears naturally if neutrinos have nonvanishing
masses.

The neutrino masses imply at least a minimal extension of the SM, 
where a mismatch
in diagonalizing mass matrices of charged and neutral leptons results
in a $3\times 3$ flavour mixing matrix. It enters the leptonic charged
current interactions, that generalizes
the interaction (\ref{ham}) to the \emph{nondiagonal}
$\nu_H$ $\leftrightarrow$ $\nu_L$ transition
\begin{equation}
\mathcal{H}^{\text{non-diag}}_\text{rad}(x)=
\frac{G_F}{\sqrt{2}}\,\bar{e}\gamma^\mu (f_V - f_A\gamma_5)
e\, \bar{\nu}_L \gamma_{\mu}(1-\gamma_5)\nu_H \;.
\label{hamnd}
\end{equation}
Here, $\nu_H$ and $\nu_L$ are just generic labels for neutrino
mass eigenstates.
The form of Eq. (\ref{hamnd}) explicates the lepton conversion 
in the neutrino sector, rather than in the charged lepton sector.

In a more ambitious approach, where one attempts to explain neutrino
masses, one faces the theories that
include non-standard interactions (NSI) of neutrinos with
matter \cite{HuSV02}. In particular,  we focus here to the neutrino
conversion via left- ($L=(1-\gamma_5)/2$) and right-handed ($R=(1+\gamma_5)/2$)
neutral current NSI of the form
\begin{equation}
\mathcal{H}_{\text{eff}}^{\text{NSI}} = \sqrt{2} G_F
\big[ \epsilon^{eL}_{\alpha\beta} \bar{e} \gamma^{\mu} L e +
      \epsilon^{eR}_{\alpha\beta} \bar{e} \gamma^{\mu} R e \big]
  \big[ \bar{\nu}_\alpha \gamma_\mu (1-\gamma_5) \nu_\beta \big]
\label{hnsi}
\end{equation}
explored further by \cite{BeR01,DaPRS03}.
Such neutral current interactions may be confused
with the higher loop effects of the charged currents, comprised
in the $f_{V,A}$ terms in (\ref{hamnd}).

In the next section we will present a novel contribution to the
axial form-factor $f_A$ in (\ref{hamnd}) arising from the charged current
transition at the two-loop level.
In Sect. 3 we will demonstrate that, due to the fact that the
SM based loop contribution is small, there is ample space left for a
discovery of certain NSI contributions.

\section{The Neutrino Conversion in the Presence of the
Axial Coupling to Electrons}

\begin{figure}
\centerline{\includegraphics[scale=0.74]{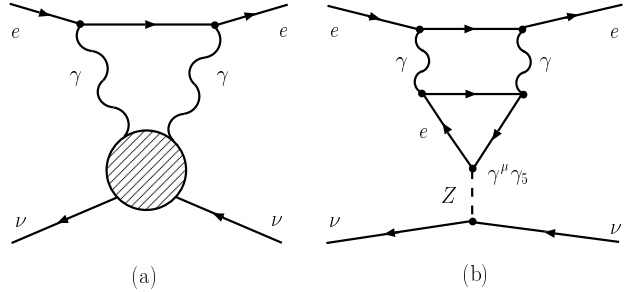}}
\vspace*{3ex}
\caption{\small Radiative corrections of order $\alpha^2$ displayed in
the 1PI blob (a), contain a nonrenormalizable triangle diagram
in the axial-vector part of the neutral current amplitude (b).
\label{triangle}}
\end{figure}

Let us recall a heuristic role of an early study of radiative corrections
to the neutrino--electron scattering. The radiative correction
at order $\alpha^2$ displayed on Fig.~\ref{triangle} generates
an infinite nonrenormalizable contribution to the $g_A$ form-factor
of the diagonal interaction (\ref{ham}).
Adler \cite{Ad69} observed this infinite radiative 
correction to $\nu_{l}l$ scattering
by embedding the triangle diagram calculated by Rosenberg \cite{Ro63}
into the next loop. In present day terminology,
the axial triangle anomaly resides in the neutral current piece of the
$\nu$--$e$ interaction displayed on Fig.~\ref{triangle}(b).

Among possible remedies for infinities proposed by Adler \cite{Ad69}, 
the idea to add
to the electron triangle on Fig.~\ref{triangle}(b) the muon triangle
with opposite sign, survived in its essence to the present day.
The quantum numbers assigned to the SM fermion
representations confirmed such cancellation for each generation of fundamental
fermions, and this \emph{anomaly cancelation} remained as a guiding
principle for model building ever since.

In addition to this piece of the neutrino axial current for which nature
took care by itself, neutrino masses invoke another contribution at
same order of $\alpha^2$, that is finite.
Up to our knowledge, this extra part to which we turn in this paper,
is novel in the literature.

Let us look how such counterpart to original Adler's contribution
sets in for massive neutrinos.
In this case, the neutrino flavour eigenstates are mixtures
of the mass eigenstates, described by the leptonic 3$\times$3 unitary matrix
(nowadays dubbed the Pontecorvo-Maki-Nakagawa-Sakata $U_{\text{PMNS}}$ matrix 
\cite{Po58,MaNS62}, in analogy to the
Cabibbo-Kobayashi-Maskawa matrix in the quark sector of the Standard model)
\begin{equation}
\left(\begin{array}{c}
\nu_e  \\ \nu_{\mu}  \\ \nu_{\tau}
\end{array}\right)
=
\left(\begin{array}{ccc}
U_{e1} & U_{e2}  & U_{e3} \\
U_{\mu 1} & U_{\mu 2}  & U_{\mu 3} \\
U_{\tau 1} & U_{\tau 2}  & U_{\tau 3}
\end{array}\right)
\left(\begin{array}{c}
\nu_1  \\ \nu_{2}  \\ \nu_{3}
\end{array}\right)
\equiv
U_{\text{PMNS}}
\left(\begin{array}{c}
\nu_1  \\ \nu_{2}  \\ \nu_{3}
\end{array}\right) \;.
\end{equation}
Note that the flavour violation 
induced by the PMNS mixing (similarly to the one by the CKM mixing) 
does not affect the renormalizability of the electroweak theory.
Thus, a safe evaluation of the quantum-loop corrections is possible.
Recently, we employed this framework in calculating lepton-flavour
violating annihilation of muonium \cite{EeKP00,EeKP01}.
Now we consider the radiative corrections which will allow for the
neutrino conversion via
the two-photon exchange similar to Fig.~\ref{triangle}(a).

\begin{figure}
\centerline{\includegraphics[scale=0.70]{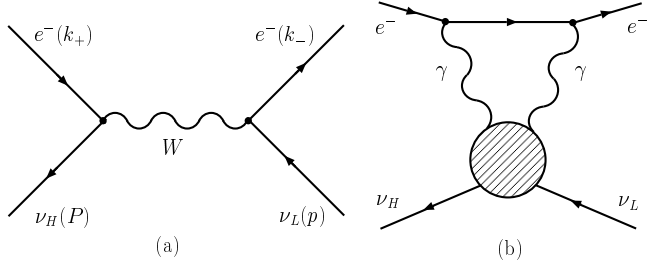}}
\caption{\small Tree-level Feynman diagram (a), describing the nondiagonal 
neutrino--electron interaction, and
the photonic-loop diagram (b) induced by the
$\nu_{H}\nu_{L}\gamma\gamma$ vertex (the shaded circle).
\label{figtree}}
\end{figure}

Since the PMNS mixing affects only the charged current, our
starting point is the tree diagram displayed on Fig.~\ref{figtree}(a).
Its two-photon radiative corrections will result in Fig.~\ref{figtree}(b) as
the nondiagonal counterpart to Fig.~\ref{triangle}.

The referent tree-level amplitude corresponding to the diagram on
Fig.~\ref{figtree}(a) reads (after a Fierz transformation)
\begin{equation}
 \mathcal{A}_{\text{tree}} = \frac{G_F}{\sqrt{2}} \sum_{\alpha=\mu,\tau}
  \lambda_\alpha \: \bar{u}(p)\gamma^{\mu}(1-\gamma_5)u(P) \:
  \bar{u}(k_-)\gamma_{\mu}(1-\gamma_5)v(k_+) \;,
\label{tree}
\end{equation}
where the summation over combinations $\lambda_\alpha \equiv 
U_{\alpha H}^{*}U_{\alpha L}$
appears on account of the unitarity of $U_{\text{PMNS}}$. 
Concerning the radiative corrections at one-loop (1L) level,
we refer to  \cite{HoKMP99} (whose notation and kinematics we keep 
on Fig~\ref{figtree}a and in our formulae for easier comparison). 
Notably, the diagram corresponding
to the one-photon variant of Fig.~\ref{figtree}b dominates
in the set of electroweak
diagrams considered in \cite{HoKMP99}.
The pertinent amplitude
\begin{equation}
 \mathcal{A}_{\text{rad}}^{\text{1L}} = \frac{G_F}{\sqrt{2}} 
  \frac{e^2}{24 \pi^2}
\left[ \sum_{\alpha=\mu,\tau} \lambda_\alpha 
\ln\frac{m_{\alpha}^2}{m_{e}^2}\right]
\bar{u}(p)\gamma^{\mu}(1-\gamma_5)u(P) \:
   \bar{u}(k_-)\gamma_{\mu}v(k_+) \;,
\label{1L}
\end{equation}
contains purely vector electron current. 
Thus, in the sum of (\ref{tree}) and (\ref{1L})
\begin{equation}
 \mathcal{A}_{\text{tree}} + \mathcal{A}_{\text{rad}}^{\text{1L}}
 =  \frac{G_F}{\sqrt{2}}
 \bar{u}(p)\gamma^{\mu}(1-\gamma_5)u(P) \:
 \bar{u}(k_-)\gamma_{\mu}(f_{V}- f_{A}\gamma_5)v(k_+) \;,
\label{tree1L}
\end{equation}
the mentioned one-loop radiative correction modifies 
only the {\em vector} form factor, and gives
\begin{equation}
  f_V  =  \sum_{\alpha=\mu,\tau} \lambda_\alpha (1+f^{\text{1L}}_\alpha) \;.
\label{gV}
\end{equation}
Thereby, the correction term in eq. (\ref{gV})  acquires a simple leading
logarithmic form,
\begin{equation}
  f^{\text{1L}}_\alpha  =  
\frac{\alpha}{3\pi}\ln\frac{m_{\alpha}}{m_{e}} \label{gV1L}\;.
\end{equation}

Now we turn to radiative corrections indicated on
Fig.~\ref{figtree}(b), which will give a contribution to the 
{\em axial-vector} form factor $f_A$ in (\ref{tree1L}).
This contribution corresponds to the two-loop electroweak diagrams
considered by us \cite{EeKP98} in the context of the fla\-vo\-ur-chan\-ging
$s\bar{d}\to\mu^+ \mu^-$ transitions.

\begin{figure}
\centerline{\includegraphics[scale=0.65]{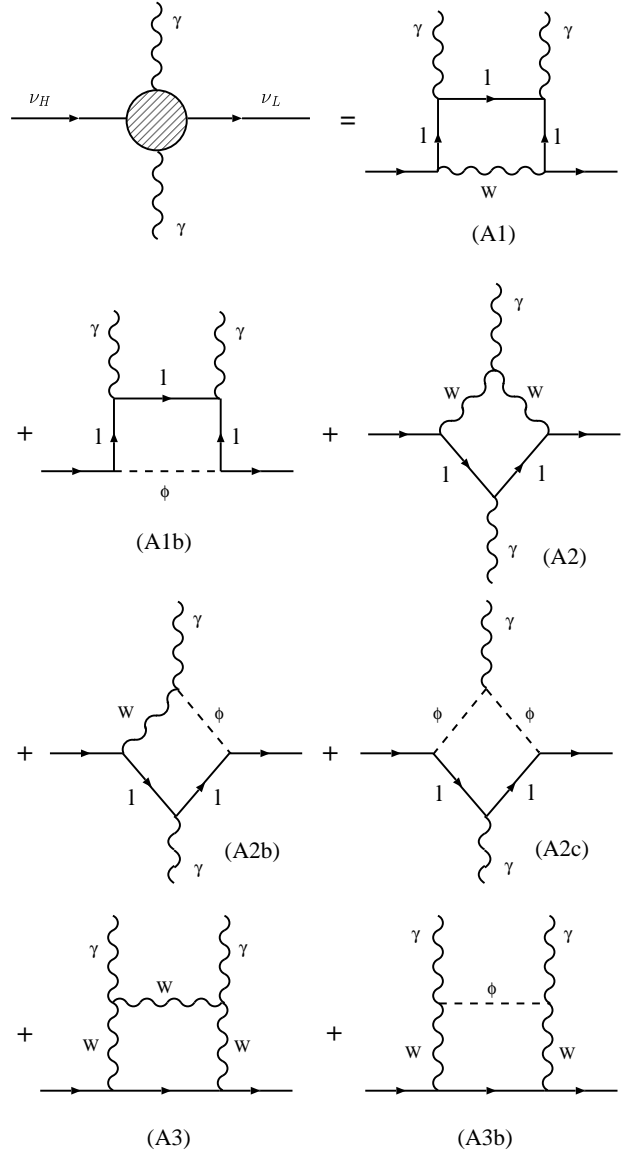}}
\caption{\small Electroweak one-loop Feynman diagrams for
$\nu_{H}\to\nu_{L}\gamma\gamma$ in the 't Hooft-Feynman gauge
 contributing to nondiagonal $\nu$--$e$ scattering on
Fig.~\protect\ref{figtree}(b).
Diagram (A2b) stands also for its counterpart diagram with $\phi$ and $W$
exchanged.
There are some additional diagrams which do not contribute
to the total $\nu_{L}\nu_{H}e e$ amplitude, as discussed in text.
\label{fig1PI}}
\end{figure}

After adding the photon-crossed counterpart to the diagram on 
Fig.~\ref{figtree}(b), the terms symmetric in indices $\mu-\nu$ cancel,
and we are left with the amplitude
\begin{equation}
\mathcal{A}^{\text{2L}}_{\text{rad}} =
-2 e^2 \bar{u}(k_-)\gamma_{\beta}\gamma_5 v(k_+)
\int \frac{d^4 q}{(2\pi)^4} \frac{1}{q^4}
\epsilon^{\beta\mu\sigma\nu} q_{\sigma} \bar{u}(p) M_{\mu\nu} u(P) \;.
\label{levicivita}
\end{equation}
Here $M_{\mu\nu}$ denotes the $\nu_{H}\nu_{L}\gamma\gamma$ vertex, and the
one-particle irreducible diagrams contributing to it in
't Hooft-Feynman gauge are displayed on Fig.~\ref{fig1PI}.
Diagrams similar to (A3) and (A3b) but with Higgs boson $\phi$
in place of one or both of the vertical-line $W$ bosons turn 
out to be $\mu-\nu$ symmetric, so they don't contribute after
contraction with Levi-Civita tensor in (\ref{levicivita}).
Diagrams employing four-boson $WW\gamma\gamma$ and $\phi\phi\gamma\gamma$
vertices vanish for the same reason.

The respective insertions
enumerated on Fig.~\ref{fig1PI} build up the two-loop
radiative amplitude
\begin{equation}
 \mathcal{A}_{\text{rad}}^{\text{2L}} = \frac{G_F}{4\sqrt{2}} 
  \frac{9}{4}\frac{\alpha^2}{\pi^2}
  \sum_{\alpha=\mu,\tau} \lambda_\alpha A_{(\alpha,e)}
\bar{u}(p)\gamma^{\mu}(1-\gamma_5)u(P) \:
   \bar{u}(k_-)\gamma_{\mu}\gamma_5 v(k_+)  \;,
\end{equation}
which modifies the axial-vector form factor $f_A$. Note that 
potentially large logarithms of the type $\ln M_{W}^{2}/m_{l}^2$ get
suppressed by the GIM mechanism. 

The net result
\begin{equation}
  f_A  =  \sum_{\alpha=\mu,\tau} \lambda_\alpha (1 + 
\frac{9 \alpha^2}{16 \pi^2}A_{(\alpha,e)}) \;, \\
\label{urgA}
\end{equation}
is expressed in terms of the GIM-like combinations $A_{(\mu,e)}$ and 
$A_{(\tau,e)}$, displayed in Table \ref{tblA}.

\begin{table}
\caption{\small Contributions $A1,\ldots,A3b$ from Fig.~\ref{fig1PI} leading
to the pertinent GIM-like loop-diagram factors $A_{(\tau,e)}$ and
$A_{(\mu,e)}$.
\label{tblA}}
\setlength{\tabcolsep}{4ex}
\renewcommand{\arraystretch}{1.2}
\centerline{\begin{tabular}{cD{.}{.}{-1}D{.}{.}{-1}}
\\
Diagram  & \multicolumn{1}{c}{$A_{(\tau,e)}$ } &
   \multicolumn{1}{c}{$A_{(\mu,e)}$}\\
\hline
  A1    &  21.4     &    14.0     \\
  A1b   &  -0.002   &    \sim 0     \\
  A2    &  0.0007   &    \sim 0     \\
  A2b    &  0.12   &     0.0007     \\
  A2c    &  0.009   &    \sim 0     \\
  A3    &  0.4   &       0.0001     \\
  A3b    &  0.03   &    \sim 0     \\
\hline 
Total   &  21.6     &   14.0 \\
\end{tabular}}
\end{table}

Since the dominant contribution comes from the (A1) diagram
in Fig.~\ref{fig1PI}, one can rely on the simple leading-log
analytical form of these
functions ($A_{(\alpha,e)}\propto f_{\alpha}^{\text{2L}}$) in
correspondence to the analytical expressions deduced previously 
\cite{VoS76,EeKP98}. In close analogy to the one-loop radiative
correction in (\ref{gV}) and (\ref{gV1L}), our two-loop radiative
correction reads
\begin{eqnarray}
  f_A & = & \sum_{\alpha=\mu,\tau} \lambda_\alpha 
(1+f^{\text{2L}}_\alpha) \;, \label{gA}\\
  f^{\text{2L}}_\alpha & = & \frac{3}{4}\frac{\alpha^2}{\pi^2}
\ln\frac{m_{\alpha}^2}{m_{e}^2} \label{gA2L}\;.
\end{eqnarray}
The dominant $\tau$-loop ($\alpha=\tau$) corrections to the referent tree-loop
amplitude, given by expressions (\ref{gV1L}) and (\ref{gA2L}), are
numerically
\begin{equation}
 f^{\text{1L}}_\tau\simeq 6.3\times 10^{-3} \;, \quad
f^{\text{2L}}_\tau \simeq 6.6\times 10^{-5} \;.
\label{1L2L}
\end{equation}
In order to estimate $f_V$ and  $f_A$, one has to include also the
PMNS-matrix prefactors, for which we now have the first strong experimental
hints.

\section{Conclusion}

The existing neutrino experiments
enable quite accurate global fits on the PMNS matrix elements. Unanticipated
outcome of these experiments are large mixings. Let us, for
definiteness, refer to the result of an earlier analysis by Fukugita
and Tanimoto \cite{FuT01}, based on the solar (SNO \cite{SNO01}) and
atmospheric (SuperKamiokande \cite{SK98}) neutrino data.
Their solution
shows a ``democratic'' mixing, very different from the ``hierarchical''
one experienced in the quark sector.
The least known matrix element has been constrained here by the
CHOOZ reactor experiment ($|U_{e3}|<0.16$).

Let us stress that for sufficiently heavy neutrino ($\nu_H$), 
such as the one allowed by the existing \emph{direct} experimental mass limit 
of $\sim$18 MeV for $\nu_{\tau}$, the diagrams in Fig.~\ref{figtree}
could give rise to the 
$\nu_{H}(P)\to\nu_{L}(p)e^{+}(k_{+})e^{-}(k_{-})$ decay.
This decay was originally considered in \cite{Sh81}, used for
constraining the $|U_{e3}|$ PMNS matrix element in \cite{Ha95},
and reconsidered later in \cite{HoKMP99}.
However, the recent measurements squeeze neutrinos to sub-eV
mass eigenstates that exclude this decay, and leave us with the
scattering variant studied here.

In the meantime, an additional possibility to constrain $|U_{e3}|$ came with
the results of KamLAND experiment \cite{kamland03}. 
For example, it allows in Ref. \cite{FuTY03}
to deduce the bounds $0.04 < |U_{e3}| < 0.19$.
This prediction came on account of assuming Fritzsch-type lepton
mass matrices adopted earlier by the same authors \cite{FuTY93}. 
The range of parameters
allowed by the KamLAND, as given by \cite{FuTY03}
\begin{equation}
\setlength{\arraycolsep}{6pt}
 U_{\text{PMNS}}= \left(
\begin{array}{ccc}
0.76-0.86 & 0.50-0.63  & 0.04-0.19 \\
0.27-0.48 & 0.63-0.72  & 0.60-0.71 \\
0.34-0.49 & 0.42-0.58  & 0.71-0.80 \\
\end{array}\right) \:,
\label{PMNS}
\end{equation}
turns out to be quite narrow and is a subject of future tests in
neutrino experiments.
These values enable us to give a definite prediction for $f_{V,A}$, which
can be compared with the corresponding NSI contributions in (\ref{hnsi}).

The present
bounds for the relevant NSI parameters are quite modest \cite{DaPRS03}
\begin{equation}
  |\epsilon^{eL}_{\tau e}| < 0.4\;, \qquad |\epsilon^{eR}_{\tau e}| < 0.7 \;.
\label{bounds}
\end{equation}
They compare to $f_{L,R}= (f_{V}\pm f_{A})/2$ combinations of form
factors (\ref{gV}) and (\ref{gA})  
\begin{equation}
 |f_L| < 0.16\;, \qquad |f_R| < 1.3 \times 10^{-3} \;,
\end{equation}
where the loop results in (\ref{1L2L}) are further
modulated by the matrix elements from (\ref{PMNS}). Thereby,
the leading contribution given by $f^{1L}_{\tau}$ in (\ref{1L2L})
is numerically further suppressed.
Consequently, the ``standard'' radiative corrections encoded
in (\ref{hamnd}) can be confused with the left-handed
$\epsilon^{L}$-couplings in (\ref{hnsi}), whereas the radiative
corrections are small enough to allow for the discovery of the
right-handed non-standard neutrino interactions.

\end{document}